\documentclass{aa}
\usepackage{natbib}
\usepackage{txfonts}
\usepackage{hyperref}
\usepackage{graphicx}
\graphicspath{{figures/}}

\begin{document}
 
\title{Anisotropic turbulent transport with horizontal shear\\ in stellar radiative zones}

\author{V. Prat \and S. Mathis}

\institute{
    AIM, CEA, CNRS, Universit\'e Paris-Saclay, Universit\'e Paris Diderot, Sorbonne Paris Cit\'e, F-91191 Gif-sur-Yvette, France
}

\date{}

\abstract
{Turbulent transport in stellar radiative zones is a key ingredient of stellar evolution theory, but the anisotropy of the transport due to the stable stratification and the rotation of these regions is poorly understood.
The assumption of shellular rotation, which is a cornerstone of the so-called rotational mixing, relies on an efficient horizontal transport.
However, this transport is included in many stellar evolution codes through phenomenological models that have never been tested.}
{We investigate the impact of horizontal shear on the anisotropy of turbulent transport.}
{We used a relaxation approximation (also known as $\tau$ approximation) to describe the anisotropising effect of stratification, rotation, and shear on a background turbulent flow by computing velocity correlations.}
{We obtain new theoretical scalings for velocity correlations that include the effect of horizontal shear.
These scalings show an enhancement of turbulent motions, which would lead to a more efficient transport of chemicals and angular momentum, in better agreement with helio- and asteroseismic observations of rotation in the whole Hertzsprung-Russell diagram.
Moreover, we propose a new choice for the non-linear time used in the relaxation approximation, which characterises the source of the turbulence.}
{For the first time, we describe the effect of stratification, rotation, and vertical and horizontal shear on the anisotropy of turbulent transport in stellar radiative zones.
The new prescriptions need to be implemented in stellar evolution calculations.
To do so, it may be necessary to implement non-diffusive transport.}


\maketitle

\section{Introduction}

The transport of angular momentum and chemical species plays a key role in stellar evolution.
In particular, rotation induces transport both through large-scale motions such as meridional circulation and through turbulent, small-scale motions generated by hydrodynamical instabilities \citep{Zahn83, Zahn92, MaederZahn98, MathisZahn04, Maeder09}.
These multi-dimensional small-scale transport processes cannot be resolved in one-dimensional (1D) stellar evolution codes, so their secular effects have to be modelled in those codes.

In the last decades, helio- and asteroseismology have provided us with many constraints on the internal rotation of the Sun and distant stars.
In the case of the Sun, helioseismic data bring to light an almost flat rotation profile in the solar radiative zone down to $0.2\,{\rm R}_\odot$, and a possibly faster core \citep{Brown,Thompson,Garcia,Fossat}.
\citet{Benomar} detected only weak differential rotation on a few solar-type stars.
For a large number of subgiant and red giant stars, asteroseismic data show a weak core-surface rotation contrast \citep{Beck,Mosser,Deheuvels12,Deheuvels14,Deheuvels15,Triana17,Gehan}.
Weak differential rotation has also been detected in intermediate-mass and massive stars \citep{Kurtz,Saio,Triana15,Murphy,VanReeth16,Aerts,VanReeth18, Ouazzani}.
Finally, a strong transport of angular momentum is needed to explain data obtained for white dwarves \citep{Suijs,Hermes} and neutron stars \citep{Heger,HirschiMaeder}.

\citet{Zahn92} introduced a formalism for turbulent transport accounting for both advection by the meridional circulation and the anisotropic turbulent transport induced by differential rotation, which is modelled thanks to vertical and horizontal turbulent diffusion coefficients.
The anisotropy of the transport is due to the combined effect of rotation and stable stratification, which limit horizontal and vertical motions through the buoyancy force and the Coriolis acceleration, respectively.
Those coefficients are based on phenomenological arguments and thus introduce large uncertainties in stellar evolution models.
A lot of work has been done to improve the physics of the vertical diffusion coefficient by adding physical ingredients such as gradients of mean molecular weight \citep{MaederMeynet,TalonZahn}, and local numerical simulations have been performed to test existing models \citep{PratLignieres13,PratLignieres14,Garaud16} and investigate the effect of viscosity \citep{PratGVM,Garaud17, GagnierGaraud}.
Current models predict less transport than what is needed to explain observations of internal rotation \citep{Eggenberger, Ceillier, Marques, Cantiello, Ouazzani}.

Concerning the horizontal diffusion coefficient, since the inital model of \citet{Zahn92} that was based on phenomenological arguments, other models based on energetic arguments \citep{Maeder03} or results of an unstratified Taylor-Couette experiment \citep{Mathis04} have been proposed, but this coefficient is still poorly constrained, and its impact on stellar evolution is clearly overlooked.
The first numerical simulations of the turbulent transport triggered by the instabilities of a horizontal shear in stellar conditions have also been performed \citep{Cope, Garaud20}.
Finally, \citet{Park20a, Park20b} studied the linear instabilities of the horizontal differential rotation in stellar radiation zones as a function of stratification, rotation, and thermal diffusivity.

However, the widely used formalism of \citet{Zahn92} assumes that the turbulent transport of angular momentum is viscous, which is not necessarily the case.
Indeed, anisotropic turbulent motions can generate non-zero turbulent fluxes of angular momentum even in the absence of differential rotation, which is a characteristic of non-diffusive transport.
The existence of such non-diffusive turbulent transport associated with the Coriolis force, known as the $\Lambda$ effect \citep{Ruediger1989}, has been confirmed by direct numerical simulations \citep[see e.g.][]{Kapyla}.
Originally, this effect was introduced in the context of solar and stellar convective zones to explain the persistance of differential rotation despite the presence of a magnetic field, which was supposed to erase it.
\citet[hereafter referred to as KB12]{KB12} later investigated the effect of the stable stratification of stellar radiative zones on the Reynolds stress.

Based on this work, \citet[hereafter referred to as M+18]{Mathis18} proposed a new model of the horizontal transport induced by the vertical shear instability.
In particular, M+18 added a vertical shear (due to radial differential rotation) to the formalism of KB12 and show that it has no significant effect on the anisotropy of turbulence.
Besides, they used estimates of mean velocity correlations to propose new prescriptions for the horizontal turbulent diffusion coefficient of chemical elements.
The implementation of these prescriptions in stellar evolution computations shows a slightly enhanced transport of angular momentum throughout the main sequence, but not enough to fit helio- and asteroseismic observations.
Furthermore, a major caveat of this implementation is that it assumed that the transport of angular momentum had the same (diffusive) nature as the transport of chemical elements, and thus completely ignored the $\Lambda$ effect.

In the present work, we perform a new generalisation of the formalism of KB12 in the presence of a general (both vertical and horizontal) shear.
It allows us to propose new prescriptions for the transport of both angular momentum and chemical elements that include the $\Lambda$ effect.

In Sect~\ref{sec:spec}, we derive spectral properties of the flow anisotropised by stratification, rotation, and shear.
Then, in Sect.~\ref{sec:corr}, we deduce scalings for velocity correlations as a function of stratification, rotation, and shear.
We propose possible interpretations of the new scalings in Sect.~\ref{sec:inter}.
We explain how non-viscous turbulent transport of angular momentum should be implemented in stellar evolution codes to account for the $\Lambda$ effect in Sect.~\ref{sec:nonvisc}.
Finally, we conclude on the impact of these results on the modelling of turbulence in stellar evolution codes in Sect.~\ref{sec:conc}.

\section{Turbulent spectrum}
\label{sec:spec}

In the Boussinesq approximation, which is justified here because turbulent motions are at length scales much smaller than the pressure scale height and velocity scales much smaller than the sound speed, the equations governing the flow are
\begin{align}
    \vec\nabla\cdot\vec V=0,    \label{eq:continuity}\\
    \frac{\partial\vec V}{\partial t}+(\vec V\cdot\vec\nabla)\vec V = -\frac{\vec\nabla P'}{\rho}+\frac{\rho'}{\rho}\vec g+\nu\Delta\vec V+\vec f,    \label{eq:momentum}\\
    \frac{\partial s}{\partial t}+\vec V\cdot\vec\nabla s = \kappa\Delta s, \label{eq:entropy}
\end{align}
where $\vec V$ is the velocity, $P'$ and $\rho'$ are pressure and density fluctuations, respectively, $\rho$ is the background density, $\vec g$ is the gravity vector, $\nu$ is the kinematic viscosity, $\vec f$ is a forcing term, $s$ is the specific entropy, and $\kappa$ is the thermal diffusivity.
If one neglects the large-scale meridional circulation, velocity can be split into a mean rotation part and a fluctuating part:
\begin{equation}
    \vec V = r\sin\theta\Omega(r,\theta)\vec e_\varphi + \vec u,
\end{equation}
where $r$, $\theta$, and $\varphi$ are the standard spherical coordinates, $(\vec e_{\rm r}, \vec e_\theta, \vec e_\varphi)$ is the associated orthonormal basis, $\Omega$ is the rotation rate, and $\vec u$ is the velocity fluctuation.
The continuity equation~\eqref{eq:continuity} implies that
\begin{equation}
    \label{eq:cont}
    \vec\nabla\cdot\vec u = 0.
\end{equation}
The left-hand side of the momentum equation~\eqref{eq:momentum} can be developed into
\begin{equation}
    \frac{\partial\vec u}{\partial t}-s\Omega^2\vec e_{\rm s}+\Omega\frac{\partial\vec u}{\partial\varphi}+s(\vec u\cdot\vec\nabla\Omega)\vec e_\varphi+\vec\Omega\wedge\vec u+(\vec u\cdot\vec\nabla)\vec u,
\end{equation}
where $s$ is the distance to the rotation axis and $\vec e_{\rm s}$ the associated unit vector.
The second term, which is the centrifugal acceleration, will be neglected in the following.
Then, in the frame that locally rotates at the same rate as the fluid, Eq.~\eqref{eq:momentum} becomes
\begin{equation}
    \label{eq:mom_rot}
    \frac{\partial\vec u}{\partial t}+(\vec S\cdot\vec u)\vec e_\varphi+2\vec\Omega\wedge\vec u+(\vec u\cdot\vec\nabla)\vec u = -\frac{\vec\nabla P'}{\rho}+\frac{\rho'}{\rho}\vec g+\nu\Delta\vec u+\vec f,
\end{equation}
where $\vec S=r\sin\theta\vec\nabla\Omega$ is the shear rate.

Similarly, the entropy can be split into a mean part $\langle s\rangle$ and a fluctuating one $s'$:
\begin{equation}
    s=\langle s\rangle + s'.
\end{equation}
Equation~\eqref{eq:entropy} yields
\begin{equation}
    \label{eq:ent_rot}
    \frac{\partial s'}{\partial t}+\vec u\cdot\vec\nabla \langle s\rangle + \vec u\cdot\vec\nabla s' = \kappa\Delta s.
\end{equation}

Non-linear and dissipative terms are usually difficult to model analytically in turbulent flows.
To simplify the problem, we use here a relaxation approximation, also known as $\tau$ approximation, which assumes that in a stationary steady state, the main effect of these terms is to tend the flow to relax with a given time constant $\tau$.
Equations~\eqref{eq:mom_rot} and \eqref{eq:ent_rot} can then be approximated by
\begin{align}
    \frac{\vec u}{\tau}+(\vec S\cdot\vec u)\vec e_\varphi+2\vec\Omega\wedge\vec u = -\frac{\vec\nabla P'}{\rho}+\frac{\rho'}{\rho}\vec g+\vec f,   \\
    \frac{s'}{\tau}+\vec u\cdot\vec\nabla \langle s\rangle = 0.
\end{align}
For a complete treatment of thermal diffusion, we refer the reader to \citet{Park20a, Park20b}.

These two equations can be combined into a single one by expressing entropy fluctuations as a function of density fluctuations:
\begin{equation}
    s'=c_{\rm P}\frac{\rho'}{\rho},
\end{equation}
where $c_{\rm P}$ is the specific heat capacity at constant pressure.
This yields
\begin{equation}
    \label{eq:mom_rel}
    \frac{\vec u}{\tau}+(\vec S\cdot\vec u)\vec e_\varphi+2\vec\Omega\wedge\vec u = -\frac{\vec\nabla P'}{\rho}-\tau N^2u_{\rm r}\vec e_{\rm r}+\vec f,
\end{equation}
where the Brunt-V\"ais\"al\"a frequency $N$ is defined by
\begin{equation}
    N^2=-\vec g\cdot\vec\nabla\langle s\rangle / c_{\rm P}.
\end{equation}
Using the Fourier decomposition
\begin{equation}
    \vec u = \sum_{\vec k} \hat{\vec u}(\vec k)e^{i\vec k\cdot\vec r},
\end{equation}
the continuity equation~\eqref{eq:cont} becomes
\begin{equation}
    \vec k\cdot\hat{\vec u} = 0,
\end{equation}
and pressure fluctuations can be eliminated from Eq.~\eqref{eq:mom_rel}:
\begin{equation}
    \hat{\vec u}+(\tilde{\vec S}\cdot\hat{\vec u})(\vec e_\varphi-\hat{k_\varphi}\hat{\vec k})+\sigma\tilde\Omega\hat{\vec k}\wedge\hat{\vec u} +\tilde N^2\hat u_{\rm r}(\vec e_{\rm r}-\mu\hat{\vec k})= \hat{\vec u}^{(0)},
\end{equation}
where $\tilde{\vec S}=\tau\vec S$, $\hat{\vec k}=\vec k/k$, $\sigma=\hat{\vec k}\cdot\vec e_{\rm z}$, $\tilde\Omega=2\tau\Omega$, $\tilde N=\tau N$, $\mu=\hat k_{\rm r}$ and $\hat{\vec u}^{(0)}=\tau\vec f$.
This way of writing the forcing term means that we assume that it comes from a pre-existing background turbulent flow.
The previous equation can be written as a matrix relation
\begin{equation}
    \label{eq:u_u0}
    \vec{\mathcal{M}}\cdot\hat{\vec u}=\hat{\vec u}^{(0)},
\end{equation}
where $\vec{\mathcal{M}}$ is the matrix
\begin{equation}
    \begin{bmatrix}
        1-\tilde S_{\rm r}\mu\hat k_\varphi+\tilde N^2(1-\mu^2)                                                   &   -\tilde S_\theta\mu\hat k_\varphi-\sigma\tilde\Omega\hat k_\varphi  &   \sigma\tilde\Omega\hat k_\theta \\
        -\tilde S_{\rm r}\hat k_\theta\hat k_\varphi+\sigma\tilde\Omega\hat k_\varphi-\tilde N^2\mu\hat k_\theta  &   1-\tilde S_\theta\hat k_\theta\hat k_\varphi                        &   -\sigma\tilde\Omega\mu          \\
        \tilde S_{\rm r}(1-\hat k_\varphi^2)-\sigma\tilde\Omega\hat k_\theta-\tilde N^2\mu\hat k_\varphi          &   \tilde S_\theta(1-\hat k_\varphi^2)+\sigma\tilde\Omega\mu           &   1
    \end{bmatrix},
\end{equation}
and $\tilde{\vec S}=\tilde S_{\rm r}\vec e_{\rm r}+\tilde S_\theta\vec e_\theta$.

When $\vec{\mathcal{M}}$ is inversible, Eq.~\eqref{eq:u_u0} implies
\begin{equation}
    \hat{\vec u} = \vec{\mathcal{M}}^{-1}\cdot\hat{\vec u}^{(0)}.
\end{equation}
Using the fact that $\hat{\vec u}^{(0)}$ also verifies the continuity equation, the previous equation reduces to
\begin{equation}
    \hat{\vec u} = \frac{\vec{\mathcal{D}}}{A}\cdot\hat{\vec u}^{(0)},
\end{equation}
where
\begin{equation}
    \label{eq:det}
    A = 1 + \tilde N^2(1-\mu^2-\tilde S_\theta\hat k_\theta\hat k_\varphi) - \eta\tilde S\hat k_\varphi + \sigma\tilde\Omega(\sigma\tilde\Omega+\mu\tilde S_\theta-\hat k_\theta\tilde S_{\rm r})
\end{equation}
is the determinant of $\vec{\mathcal{M}}$, $\eta=\hat k\cdot \tilde{\vec S}/\tilde S$, $\vec{\mathcal{D}}$ is the matrix given by
\begin{equation}
    \begin{aligned}
        \vec{\mathcal{D}}_{ij}&=[1+\tilde N^2(1-\mu^2)]\delta_{ij}+\tilde N^2(\mu\hat k_i-\delta_{i{\rm r}})\delta_{j{\rm r}}+\sigma\tilde\Omega\varepsilon_{ijl}\hat k_l   \\
        &\quad-\eta\tilde S\hat k_\varphi\delta_{i\varphi}\delta_{j\varphi}-S_j(1-\hat k_\varphi^2)\delta_{i\varphi}-\tilde S_l\varepsilon_{il\varphi}\varepsilon_{jm\varphi}\hat k_m\hat k_\varphi,
    \end{aligned}
\end{equation}
and $\delta_{ij}$ and $\varepsilon_{ijk}$ are the Kronecker and Levi-Civita tensors, respectively.

The spectrum of the background turbulence is assumed to be characterised by the correlation tensor $Q_{ij}^{(0)}=\langle u_i^{(0)}u_j^{(0)}\rangle$, where $\langle\rangle$ denotes a statistical average.
In the continuity of \citet{Mathis18}, we assume a purely horizontal background turbulence, for which the Fourier transform of $Q_{ij}^{(0)}$ is
\begin{equation}
    \hat Q_{ij}^{(0)}=\frac{3E(k)}{8\pi k^2}[(1-\mu^2)(\delta_{ij}-\hat k_i\hat k_j)-(\delta_{i{\rm r}}-\mu\hat k_i)(\delta_{j{\rm r}}-\mu\hat k_j)],
\end{equation}
where $E(k)$ is the kinetic energy spectrum \citep[see e.g.][]{Davidson} such that
\begin{equation}
    \int_0^{+\infty} E(k){\rm d}k=\langle {\vec u}^2\rangle^{(0)}.
\end{equation}
The turbulent spectrum of the sheared turbulence is deduced from that of the background turbulence using the relation
\begin{equation}
    \hat Q_{ij}=\frac{1}{A^2}\vec{\mathcal{D}}_{im}\vec{\mathcal{D}}_{jn}\hat Q_{mn}^{(0)}.
\end{equation}
After some algebra, this leads to 
\begin{equation}
    \hat Q_{ij}=\frac{3E(k)}{8\pi k^2A^2}C_iC_j,
\end{equation}
where
\begin{align}
    C_{\rm r}   &=  \sigma\tilde\Omega(1-\mu^2)+\tilde S_\theta\mu\hat k_\varphi^2,   \\
    C_\theta    &=  [1+\tilde N^2(1-\mu^2)-\mu\hat k_\varphi\tilde S_{\rm r}]\hat k_\varphi-\sigma\tilde\Omega\mu\hat k_\theta,  \\
    C_\varphi   &=  -[1+\tilde N^2(1-\mu^2)-\mu\hat k_\varphi\tilde S_{\rm r}]\hat k_\theta-\mu\hat k_\varphi(\sigma\tilde\Omega+\mu\tilde S_\theta).
\end{align}

\section{Mean correlations}
\label{sec:corr}

Mean velocity correlations can be computed using the relation
\begin{equation}
    Q_{ij}=\int\hat Q_{ij}{\rm d}^3k.
\end{equation}

Stellar radiative zones are usually strongly stratified.
It implies that $\tilde N\gg\tilde\Omega$ \citep[for example, the typical value of the ratio $N/(2\Omega)$ is around 500 for the Sun; see also][for the evolution of this ratio along stellar evolution for low- and intermediate-mass stars]{Andre18, Andre19}, $\tilde N\gg\tilde S$, and $\tilde N\gg1$.
As a first approximation,
\begin{equation}
    \label{eq:denom}
    A\simeq \tilde N^2(1-\mu^2-\tilde S_\theta\hat k_\theta\hat k_\varphi)=\tilde N^2\sin^2\alpha(1-\tilde S_\theta\sin\beta\cos\beta),
\end{equation}
where $\mu=\cos\alpha$, $\hat k_\theta=\sin\alpha\cos\beta$ and $\hat k_\varphi=\sin\alpha\sin\beta$ (with these notations, $\sigma=\cos\theta\cos\alpha-\sin\theta\sin\alpha\cos\beta$).
For the same reason,
\begin{equation}
    \label{eq:qrt}
    \hat Q_{{\rm r}\theta} \simeq \frac{3E(k)}{8\pi k^2}\frac{\sigma\tilde\Omega\sin\alpha\sin\beta+\tilde S_\theta\sin\alpha\cos\alpha\sin^3\beta}{\tilde N^2(1-\tilde S_\theta\sin\beta\cos\beta)^2}.
\end{equation}
Using parity arguments and integrating over $\vec k$ (see Appendix~\ref{sec:integrals}), this leads, when $|\tilde S_\theta|<2$, to
\begin{equation}
    \label{eq:urut}
    \langle u_{\rm r} u_\theta\rangle\simeq-\frac{\tilde\Omega}{\tilde N^2}\frac{\tilde S_\theta\sin\theta}{(1-\tilde S_\theta^2/4)^{3/2}}\frac{\langle{\vec u^2\rangle^{(0)}}}{4}.
\end{equation}
This term corresponds to the Reynolds stress in the meridional plane.
Similarly,
\begin{align}
    \langle u_{\rm r} u_\varphi\rangle        &\simeq \frac{\tilde\Omega}{\tilde N^2}\frac{\sin\theta}{(1-\tilde S_\theta^2/4)^{3/2}}\frac{\langle{\vec u^2\rangle^{(0)}}}{2},    \label{eq:urup} \\
    \langle u_\theta u_\varphi\rangle   &\simeq -\frac{\tilde S_\theta}{(1-\tilde S_\theta^2/4)^{3/2}}\frac{\langle{\vec u^2\rangle^{(0)}}}{4}, \label{eq:utup} \\
    \langle u_\theta^2\rangle           &\simeq \frac{1}{(1-\tilde S_\theta^2/4)^{3/2}}\frac{\langle{\vec u^2\rangle^{(0)}}}{2} \simeq \langle u_\varphi^2\rangle.  \label{eq:ut2}
\end{align}
The first two terms correspond to the vertical and horizontal transport of angular momentum, respectively.
The other two describe the horizontal turbulent velocity scales.

We note that when $\tilde S_\theta$ tends towards zero, the scalings for $\langle u_{\rm r}u_\theta\rangle$ and $\langle u_\theta u_\varphi\rangle$ (in Eqs.~\eqref{eq:urut} and~\eqref{eq:utup}, respectively) vanish, while the others tend towards finite values.
These values, as well as the zero limit of $\langle u_\theta u_\varphi\rangle$, are consistent with the results of KB12.
In contrast, KB12 found a non-zero value for $\langle u_{\rm r}u_\theta\rangle$:
\begin{equation}
    \langle u_{\rm r} u_\theta\rangle_{\rm KB12} \simeq \frac{\tilde\Omega^2}{\tilde N^4}\sin\theta\cos\theta\frac{\langle{\vec u^2\rangle^{(0)}}}{2},
\end{equation}
which is of lower order in $\tilde N$ than the new scaling of Eq.~\eqref{eq:urut}.

The case of $\langle u_{\rm r}^2\rangle$ is more complicated, since it contains three terms that come from
\begin{equation}
    C_{\rm r}^2=\sigma^2\tilde\Omega^2(1-\mu^2)^2+2\sigma\tilde\Omega\tilde S_\theta(1-\mu^2)\mu\hat k_\varphi^2+\tilde S_\theta^2\mu^2\hat k_\varphi^4
\end{equation}
and a priori have the same order of magnitude.
The first term leads to
\begin{equation}
    \label{eq:ur21}
    \langle u_{\rm r}^2\rangle_1  \simeq  \frac{\tilde\Omega^2}{\tilde N^4}\frac{1}{(1-\tilde S_\theta^2/4)^{3/2}}\frac{\langle{\vec u^2\rangle^{(0)}}}{2},
\end{equation}
which is consistent with the scaling derived in KB12 and M+18.
The second term vanishes for parity reasons.
The third term leads to
\begin{equation}
    \label{eq:ur22}
    \langle u_{\rm r}^2\rangle_2 \simeq \frac{1}{\tilde N^4}\left[\frac{1}{(1-\tilde S_\theta^2/4)^{3/2}} - 1\right]\frac{\langle{\vec u^2\rangle^{(0)}}}{2}.
\end{equation}
When $|\tilde S_\theta|\ll1$, this reduces to
\begin{equation}
    \langle u_{\rm r}^2\rangle_2 \simeq \frac{3\tilde S_\theta^2}{16\tilde N^4}\langle\vec u^2\rangle^{(0)},
\end{equation}
which is small compared to $\langle u_{\rm r}^2\rangle_1$.
In total,
\begin{equation}
    \langle u_{\rm r}^2\rangle \simeq \frac{1}{\tilde N^4}\left[\frac{1+\tilde\Omega^2}{(1-\tilde S_\theta^2/4)^{3/2}} - 1\right]\frac{\langle{\vec u^2\rangle^{(0)}}}{2}.
\end{equation}

\section{Interpretation}
\label{sec:inter}

Interestingly, none of the terms computed in Sect.~\ref{sec:corr} explicitely depend on the radial shear $\tilde S_{\rm r}$, although it was included in the governing equations.
In contrast, all share a common feature: they have a denominator in $(1-\tilde S_\theta^2/4)^{3/2}$.
This implies that when $|\tilde S_\theta|$ gets close to $2$, the turbulence is enhanced.
Physically, this might be related to the limit of the Rayleigh-Taylor instability.
The Rayleigh-Taylor stability criterion is
\begin{equation}
    2\Omega(2\Omega+\vec S\cdot\vec e_{\rm s})>0.
\end{equation}
Assuming that $\Omega>0$, this criterion can be rewritten
\begin{equation}
    \cos\theta S_\theta > -(2\Omega+\sin\theta S_{\rm r}).
\end{equation}
When $2\Omega+\sin\theta S_{\rm r}>0$, the last inequality is verified for all positive values of $\cos\theta S_\theta$, but only for negative values verifying
\begin{equation}
    |S_\theta| < \left|\frac{2\Omega+\sin\theta S_{\rm r}}{\cos\theta}\right|.
\end{equation}
By comparison with $|\tilde S_\theta|<2$, this suggests that
\begin{equation}
    \tau = \left|\frac{2\cos\theta}{2\Omega+\sin\theta S_{\rm r}}\right|.
\end{equation}
This expression is similar to $\tau=(2\Omega+S_{\rm r})^{-1}$, proposed by M+18.
However, the presence of $\cos\theta$ at the numerator means that $\tau$ would vanish at the equator, in which case the scalings derived in Sect.~\ref{sec:corr} are not valid there.
Besides, the comparison with the Rayleigh-Taylor instability stands only in the special case where $2\Omega+\sin\theta S_{\rm r}>0$ and $\cos S_\theta<0$.
Thus, the criterion $|\tilde S_\theta|<2$ could be linked to some other hydrodynamical instability instead, such as the inertial or the inflectional instability studied by \citet{Park20a, Park20b}.

The form of the expression of $\langle u_\theta u_\varphi\rangle$ in Eq.~\eqref{eq:utup}, which describes the horizontal transport of angular momentum, suggests that this flux term is viscous, with a turbulent viscosity coefficient equal to
\begin{equation}
    \label{eq:nuh}
    \nu_{\rm h} = \frac{\tau\langle\vec u^2\rangle^{(0)}}{4(1-\tilde S_\theta^2/4)^{3/2}}.
\end{equation}
We note that this coefficient has no explicit dependence on the Brunt-V\"ais\"al\"a frequency or on the vertical shear.
However, $\tau$ most likely depends on $S_{\rm r}$, and not necessarily on $N$.
Thus, the viscosity coefficient probably depends on both $S_{\rm r}$ and $S_\theta$.
The dependence on $\langle\vec u^2\rangle^{(0)}$ is problematic for the use in stellar evolution codes, since there is no easy way to reliably prescribe this quantity.
One way to roughly estimate it is to assume that horizontal motions generated by the background turbulence are global.
This can be written
\begin{equation}
    \label{eq:approx}
    \langle \vec u^2\rangle^{(0)} \sim R^2 / \tau^2,
\end{equation}
where $R$ is the radial extent of the radiative zone.
Introducing this expression in Eq.~\eqref{eq:nuh} yields
\begin{equation}
    \nu_{\rm h} \simeq \frac{R^2}{4\tau(1-\tilde S_\theta^2/4)^{3/2}}.
\end{equation}
In the case where the dynamics is dominated by the horizontal shear, assuming $\tau\sim1/S_\theta$ further leads to a horizontal viscosity coefficient that is proportional to the horizontal shear rate, which is consistent with the viscosity coefficient derived by \citet{Park20b}.
We note that, in contrast, \citet{Garaud01} found that the horizontal shear instability can lead to an anti-diffusive horizontal transport of angular momentum.

Although the expression in Eq.~\eqref{eq:nuh} is not directly usable in stellar evolution codes, it can be used to express other flux terms as a function of $\nu_{\rm h}$.
Thus,
\begin{align}
    \frac{\langle u_{\rm r}u_\theta\rangle}{\nu_{\rm h}}    &\simeq -\frac{2\Omega}{\tau N^2}S_\theta\sin\theta,    \\
    \frac{\langle u_{\rm r}u_\varphi\rangle}{\nu_{\rm h}}   &\simeq \frac{4\Omega}{\tau^2N^2}\sin\theta.
\end{align}

In the present work, we consider the case of a horizontally isotropic turbulence (i.e. with identical statistical properties in the latitudinal and azimuthal directions).
This is rigourously justified for the radiative zone of slowly rotating stars, in which turbulent structures would have the shape of a horizontal pancake (see e.g. Fig.~1 of M+18).
Because of that, there is no horizontal $\Lambda$ effect.
In rapidly rotating stars, turbulent structures are elongated along the rotation axis, so the horizontal isotropy is broken \citep[see e.g.][]{Ruediger1980} and a non-zero horizontal flux of angular momentum due to the $\Lambda$ effect is present.
However, in stellar radiative zones, provided that the Brunt-V\"ais\"al\"a frequency is still much larger than the rotation rate, the horizontal flux due to the $\Lambda$ effect remains negligle compared to the leading diffusive term, which reduces to Eq.~\eqref{eq:nuh} in the horizontally symmetric case.

In contrast with the horizontal transport, the vertical transport may not be viscous, since the term $\langle u_{\rm r}u_\varphi\rangle$ given in Eq.~\eqref{eq:urup} is not explicitly proportional to $S_{\rm r}$.
Instead, it is proportional to the rotation rate $\Omega$, which is a characteristic of the $\Lambda$ effect.
However, when the dynamics is dominated by the vertical shear, one can assume that $\tau\sim 1/S_{\rm r}$, in which case Eq.~\eqref{eq:urup} becomes
\begin{equation}
    \langle u_{\rm r} u_\varphi\rangle \simeq \frac{\Omega S_{\rm r}}{N^2}\frac{\sin\theta}{(1-\tilde S_\theta^2/4)^{3/2}}\langle{\vec u^2\rangle^{(0)}}.
\end{equation}
This then leads to a vertical turbulent viscosity coefficient
\begin{equation}
    \label{eq:nuv}
    \nu_{\rm v} = -\frac{\Omega\sin\theta\langle\vec u^2\rangle^{(0)}}{N^2(1-\tilde S_\theta^2/4)^{3/2}},
\end{equation}
which is negative.
This corresponds to an anti-diffusive transport of angular momentum.
\citet{Cope} and \citet{Garaud20} estimated the vertical turbulent diffusion coefficient associated with the horizontal shear instability using scaling laws but they did not provide any proof that the transport of chemical elements is indeed diffusive.
In addition, they did not discuss the transport of angular momentum.
Using the approximation of Eq.~\eqref{eq:approx} leads to
\begin{equation}
    \nu_{\rm v}\simeq -\frac{R^2\Omega\sin\theta}{\tau^2 N^2(1-\tilde S_\theta^2/4)^{3/2}}.
\end{equation}
The same approximation can be used in the non-viscous case to express the flux from Eq.~\eqref{eq:urup}:
\begin{equation}
    \langle u_{\rm r} u_\varphi\rangle    \simeq \frac{R^2\Omega\sin\theta}{\tau^3 N^2(1-\tilde S_\theta^2/4)^{3/2}}.
\end{equation}

Following KB12 and M+18, we estimate the turbulent diffusion coefficients using a mixing-length approximation $D_{\rm v}\simeq \tau \langle u_{\rm r}^2\rangle$ and $D_{\rm h} \simeq \tau\langle u_{\rm h}^2\rangle$, where $\langle u_{\rm h}^2\rangle=\langle u_\theta^2\rangle+\langle u_\varphi^2\rangle$.
This allows us to estimate the ratio between vertical and horizontal diffusion coefficients:
\begin{equation}
    \label{eq:ratio}
    \frac{D_{\rm v}}{D_{\rm h}} \simeq \frac{\tilde\Omega^2 + 1 - (1-\tilde S_\theta^2/4)^{3/2}}{2\tilde N^4},
\end{equation}
which has the same scaling with $N$ as found by M+18.
Besides, in the limit where $\tilde S_\theta\ll1$, Eq.~\eqref{eq:ratio} reduces to
\begin{equation}
    \frac{D_{\rm v}}{D_{\rm h}} \simeq \frac{\tilde\Omega^2 + 3\tilde S_\theta^2/8}{2\tilde N^4},
\end{equation}
which also has the same scaling with $\tau$ as found by M+18.

Moreover, in contrast with M+18, where $\langle u_\theta u_\varphi\rangle$ was zero and no horizontal turbulent viscosity coefficient could be directly derived, we can now compare horizontal viscosity and diffusion coefficients.
We thus find that $D_{\rm h}\simeq4\nu_{\rm h}$.
This suggests that the common assumption that they are equal may be qualitatively valid, but given the uncertainties introduced by the approximation we used, we cannot really conclude on the quantitative validity.
In the vertical direction, we cannot directly compute the turbulent viscosity coefficient, so there is no evidence of such an equality.

\section{Non-viscous transport in stellar evolution codes}
\label{sec:nonvisc}

If a prescription of $\langle \vec u^2\rangle^{(0)}$ is found, our expressions for the flux terms in Eqs.~\eqref{eq:urup} and~\eqref{eq:utup} can be used in stellar evolution calculations.
For non-viscous transport, the evolution of angular momentum is governed by
\begin{equation}
    \label{eq:transport}
    \begin{aligned}
        &\frac{\partial}{\partial t}(\rho r^2\sin^2\theta\Omega) + \vec\nabla\cdot(\rho r^2\sin^2\theta\Omega\vec u) \\
        &\quad=-\frac{\sin\theta}{r^2}\frac{\partial}{\partial r}(\rho r^3\langle u_{\rm r}u_\varphi\rangle)-\frac{1}{\sin\theta}\frac{\partial}{\partial\theta}(\rho\sin^2\theta\langle u_\theta u_\varphi\rangle),
    \end{aligned}
\end{equation}
where $\vec u(r,\theta)$ is the large-scale velocity field of the meridional circulation in stellar radiative zones.
1D stellar evolution codes that take rotation into account usually are based on the shellular approximation, which assumes that $\Omega=\bar\Omega(r)+\hat\Omega(r,\theta)$, where
\begin{equation}
    \bar\Omega=\frac{\int_0^\pi\Omega\sin^3\theta{\rm d}\theta}{\int_0^\pi\sin^3\theta{\rm d}\theta},
\end{equation}
and $\hat\Omega\ll\bar\Omega$ \citep[see e.g.][]{Zahn92}.
In this approximation, the meridional circulation reads
\begin{equation}
    \vec u(r,\theta)=U(r)P_2(\cos\theta)\vec e_{\rm r} + V(r){\rm d}[P_2(\cos\theta)]/{ {\rm d}\theta}\vec e_\theta,
\end{equation}
where $P_2$ is the Legendre polynomial of degree 2.
Averaging Eq.~\eqref{eq:transport} over isobars yields
\begin{equation}
    \frac{\partial}{\partial t}(\rho r^2\bar\Omega) = \frac{1}{5r}\frac{\partial}{\partial r}(\rho r^4\bar\Omega U) -\frac{1}{r^2}\frac{\partial}{\partial r}(\rho r^3\overline{\langle u_{\rm r}u_\varphi\rangle}),
\end{equation}
where
\begin{equation}
    \overline{\langle u_{\rm r}u_\varphi\rangle}=\frac{\int_0^\pi\langle u_{\rm r}u_\varphi\rangle\sin^2\theta{\rm d}\theta}{\int_0^\pi\sin^3\theta{\rm d}\theta}.
\end{equation}
We note that the term of horizontal transport is no longer present since it has a zero average.
We showed in Sect.~\ref{sec:inter} that the horizontal transport can be seen as viscous.
Therefore, the equation for the fluctuations $\hat\Omega$ is the same as in \citet{Mathis04}, for example:
\begin{equation}
    \frac{\partial}{\partial t}(\rho r^2\Omega_2) - 2\rho\bar\Omega r(2V-\alpha U) = - 10\rho\nu_{\rm h}\Omega_2,
\end{equation}
where $\hat\Omega(r,\theta)=\Omega_2(r)[P_2(\cos\theta) + 1/5]$ and
\begin{equation}
    \alpha=\frac{1}{2}\frac{{\rm d}\ln r^2\bar\Omega}{{\rm d}\ln r}.
\end{equation}

\section{Conclusion}
\label{sec:conc}

In the current paper, we investigated the impact of horizontal shear on the anisotropy of the transport in stably stratified, rotating stellar radiative zones.
We were able to derive new scalings for mean velocity correlations that confirmed that the main effect is to significantly enhance turbulent motions.
Thus, horizontal shear can be considered as an additional source of transport of angular momentum and chemical elements both in the horizontal and vertical directions.

In the regime where the stratification has a much stronger effect than rotation, we found that the horizontal transport of angular momentum is diffusive, while the vertical transport is dominated by the $\Lambda$ effect.
Additionally, the horizontal turbulent viscosity and diffusion coefficients are proportional to each other, which partly justifies the approximation that they are equal, well-spread in stellar evolution codes.
The expressions of all the mean Reynolds stresses computed in the current paper depend on a property of the background turbulent flow, which is unknown a priori.
Nevertheless, if one of the components is known by another way (for example using a phenomenological model for the horizontal turbulent viscosity coefficient), this allows one to compute fluxes of angular momentum.
We finally show how the non-viscous transport of angular momentum predicted by our model should be implemented in stellar evolution codes to go beyond the current diffusive and viscous formalism.

Although the relaxation approximation has been validated in special cases by numerical simulations (see e.g. KB12), its validity in the configuration considered in the present work is uncertain.
In particular, the dependence of the non-linear time $\tau$ on the physical parameters is undetermined.
In M+18, three different physically motivated but phenomenological expressions were proposed.
Here, we derived from more rigourous arguments a new expression for the non-linear turbulent time scale which is close to one of those proposed in M+18, $\tau=(2\Omega+S_{\rm r})^{-1}$, which corresponds to a turbulence dominated by the effects of rotation and vertical shear.
Ultimately, direct numerical simulations should be performed to test the formalism and the new prescriptions in the presence of a general shear.

The results presented in this work have been obtained in the Boussinesq approximation, which neglects density fluctuations except in the buoyancy term.
This treatment is valid as long as turbulent structures are small compared to the density scale height.
When this condition is not satisfied, for example in regions with strong density gradients, similar calculations should be performed in the more general framework of the anelastic approximation, which accounts for density gradients, but filters pressure waves out.

\begin{acknowledgements}
    V.P. and S.M. acknowledge support from the European Research Council through ERC grant SPIRE 647383 and from the CNES PLATO/GOLF grant at CEA-Saclay.
    The authors thank the anonymous referee for their constructive comments, which have helped us improve the manuscript.
\end{acknowledgements}

\bibliographystyle{aa}
\bibliography{refs}

\appendix
\onecolumn

\section{Integrals}
\label{sec:integrals}

The derivation of Eq.~\eqref{eq:urut} from Eq.~\eqref{eq:qrt} requires the computation of the double integral
\begin{equation}
\int_0^\pi\int_0^{2\pi}\frac{\tilde\Omega\cos\theta\sin\alpha\cos\alpha\sin\beta-\tilde\Omega\sin\theta\sin^2\alpha\sin\beta\cos\beta+\tilde S_\theta\sin\alpha\cos\alpha\sin^3\beta}{(1-\tilde S_\theta\sin\beta\cos\beta)^2}\sin\alpha{\rm d}\alpha{\rm d}\beta.
\end{equation}
The first and third terms of the integrand are antisymmetric with respect to $\alpha=\pi/2$, so the corresponding contribution to the integral is zero.
The remaining term is then proportional to
\begin{equation}
    I_1 = \int_0^{2\pi}\frac{\sin\beta\cos\beta{\rm d}\beta}{(1-\tilde S_\theta\sin\beta\cos\beta)^2} = \frac12\int_0^{2\pi}\frac{\sin2\beta{\rm d}\beta}{(1-\tilde S_\theta\sin2\beta/2)^2}.
\end{equation}
Using the periodicity of the integrand and the variables $t=\tan\beta$ and $u=t-\tilde S_\theta/2$, one obtains
\begin{equation}
    I_1 = \int_{-\infty}^{+\infty}\frac{2t{\rm d}t}{(1-\tilde S_\theta t+t^2)^2}=\int_{-\infty}^{+\infty}\frac{2(u+\tilde S_\theta/2){\rm d}u}{(1-\tilde S_\theta^2/4+u^2)^2}=\tilde S_\theta\int_{-\infty}^{+\infty}\frac{{\rm d}u}{(1-\tilde S_\theta^2/4+u^2)^2}.
\end{equation}
These integrals are convergent only when $|\tilde S_\theta|<2$.
Using $v=u/\sqrt{1-\tilde S_\theta^2/4}$, the integral now reads
\begin{equation}
    I_1 = \frac{\tilde S_\theta}{(1-\tilde S_\theta^2/4)^{3/2}}\int_{-\infty}^{+\infty}\frac{{\rm d}v}{(1+v^2)^2}=\frac{\pi\tilde S_\theta}{2(1-\tilde S_\theta^2/4)^{3/2}}.
\end{equation}

Similarly, the derivation of Eq.~\eqref{eq:urup} requires the computation of
\begin{equation}
    I_2=\int_0^{2\pi}\frac{\cos^2\beta{\rm d}\beta}{(1-\tilde S_\theta\sin\beta\cos\beta)^2}=\frac{1}{2}\int_0^{2\pi}\frac{(1+\cos2\beta){\rm d}\beta}{(1-\tilde S_\theta\sin2\beta/2)^2}=2\int_{-\infty}^{+\infty}\frac{ {\rm d}t}{(1-\tilde S_\theta t+t^2)^2}=\frac{\pi}{(1-\tilde S_\theta^2/4)^{3/2}},
\end{equation}
and the derivation of Eq.~\eqref{eq:ut2} requires the computation of
\begin{equation}
    I_3=\int_0^{2\pi}\frac{\sin^2\beta{\rm d}\beta}{(1-\tilde S_\theta\sin\beta\cos\beta)^2}=\frac{1}{2}\int_0^{2\pi}\frac{(1-\cos2\beta){\rm d}\beta}{(1-\tilde S_\theta\sin2\beta/2)^2}=2\int_{-\infty}^{+\infty}\frac{t^2{\rm d}t}{(1-\tilde S_\theta t+t^2)^2}=\frac{\pi}{(1-\tilde S_\theta^2/4)^{3/2}}.
\end{equation}
It follows that
\begin{equation}
    I_4=\int_0^{2\pi}\frac{{\rm d}\beta}{(1-\tilde S_\theta\sin\beta\cos\beta)^2}=\frac{2\pi}{(1-\tilde S_\theta^2/4)^{3/2}},
\end{equation}
which is required for the derivation of Eq.~\eqref{eq:ur21}.
Finally, the derivation of Eq.~\eqref{eq:ur22} requires the computation of
\begin{equation}
    I_5=\int_0^{2\pi}\frac{\sin^4\beta{\rm d}\beta}{(1-\tilde S_\theta\sin\beta\cos\beta)^2}=\frac{1}{4}\int_0^{2\pi}\frac{(1-\cos2\beta)^2{\rm d}\beta}{(1-\tilde S_\theta\sin2\beta/2)^2}=2\int_{-\infty}^{+\infty}\frac{t^4{\rm d}t}{(1+t^2)(1-\tilde S_\theta t+t^2)^2}.
\end{equation}
Using the partial fraction decomposition
\begin{equation}
    \frac{t^4}{(1+t^2)(1-\tilde S_\theta t+t^2)^2}= - \frac{1}{\tilde S_\theta}\frac{(1-\tilde S_\theta)t+\tilde S_\theta}{(1+t^2-\tilde S_\theta t)^2} + \frac{1+\tilde S_\theta^2}{\tilde S_\theta^2}\frac{(1-\tilde S_\theta)t+\tilde S_\theta}{1+t^2-\tilde S_\theta t} -\frac{1}{\tilde S_\theta^2}\frac{1}{1+t^2},
\end{equation}
one obtains
\begin{equation}
    I_5=\frac{2\pi}{\tilde S_\theta^2}\left[\frac{\tilde S_\theta(\tilde S_\theta^2 - 3)}{4(1-\tilde S_\theta^2/4)^{3/2}} + \frac{1+\tilde S_\theta^2}{\sqrt{1-\tilde S_\theta^2/4}} - 1\right] = \frac{2\pi}{\tilde S_\theta^2}\left[\frac{1}{(1-\tilde S_\theta^2/4)^{3/2}} - 1\right].
\end{equation}

\end{document}